\documentclass{article}
\usepackage{spconf,amsmath,amsfonts,graphicx, multirow}
\usepackage{booktabs}
\usepackage{setspace, enumitem}

\title{An Attention-based Approach to Hierarchical Multi-label Music Instrument Classification}
\name{Zhi Zhong, Masato Hirano, Kazuki Shimada, Kazuya Tateishi, Shusuke Takahashi, Yuki Mitsufuji}
\address{Sony Group Corporation, Tokyo, Japan} 
\begin{document}
\ninept
\maketitle
\begin{abstract}
    Although music is typically multi-label, many works have studied hierarchical music tagging with simplified settings such as single-label data. Moreover, there lacks a framework to describe various joint training methods under the multi-label setting. In order to discuss the above topics, we introduce hierarchical multi-label music instrument classification task. The task provides a realistic setting where multi-instrument real music data is assumed. Various hierarchical methods that jointly train a DNN are summarized and explored in the context of the fusion of deep learning and conventional techniques. For the effective joint training in the multi-label setting, we propose two methods to model the connection between fine- and coarse-level tags, where one uses rule-based grouped max-pooling, the other one uses the attention mechanism obtained in a data-driven manner. Our evaluation reveals that the proposed methods have advantages over the method without joint training. In addition, the decision procedure within the proposed methods can be interpreted by visualizing attention maps or referring to fixed rules.
\end{abstract}

\section{Introduction}
\label{sec:intro}
\begin{figure*}[tb]
        \centering
        \includegraphics[width=16cm ]{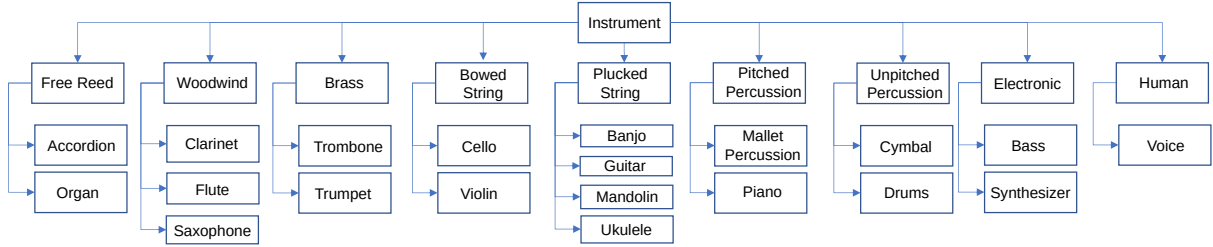} 
        \vspace{-2.5mm}
        \caption{Induced tone-base 2-level instrument hierarchy in OpenMIC dataset}
        \label{fig:OpenMIC_label}
    \end{figure*}

    Multi-label music tagging is a classification task in which the goal is to predict multiple semantic tags for a given music piece. Tags can indicate the genres, moods and instruments of the music. Therefore, this task is meaningful for applications such as music recommendation or music retrieval. Music tags organized in a tree-like structure, \textit{i.e.}, a hierarchy as shown in Fig.~\ref{fig:OpenMIC_label}, present the domain knowledge (what kind of tags are musically correlated), bringing benefits including improved tagging performance \cite{cmu2018first_hierarchy, sie2019hierarchy,sun2020graphcnn, amazon2022softtree}. While music tagging datasets typically have a flat hierarchy \cite{law2009mtat,msd2011, bogdanov2019mtgjamendo,won2020eval}, there has been growing interests in hierarchical tagging in the field of music information retrieval \cite{mir_tutorial}.
    
    Several works have tackled hierarchical music tagging. Parmezan \textit{et al.} have investigated hierarchical genre classification using conventional machine learning methods without deep learning \cite{FMAhierarchy2020}. A few works have studied deep neural network (DNN)-based hierarchical methods under a simplified problem setting. For example, Garcia \textit{et al.} tackled few-shot instrument classification for single-instrument data \cite{medlydb2021hierarchy}, while Nolasco \textit{et al.} have studied instrument representation learning with single-note data \cite{rankloss2022}. Toward hierarchical multi-label music tagging task, Krause \textit{et al.} reported on several DNN-based hierarchical methods on singing activity detection \cite{krause2022sing}. Many of their discussions are devoted to training separate DNNs, rather than training these DNNs jointly.
    
    Although real music is typically polyphonic and multi-instrument, many works have addressed hierarchical music tagging with simplified data. Therefore, DNN-based hierarchical music tagging under a realistic problem setting is yet to be discussed extensively. Moreover, there lacks a framework to describe various joint training methods under the multi-label setting. While using multiple separate models is effective, DNNs have been shown capable of learning hierarchical information during the joint training \cite{cmu2018first_hierarchy, sie2019hierarchy,sun2020graphcnn, amazon2022softtree,medlydb2021hierarchy,rankloss2022}.

    In this paper, we address hierarchical multi-label music tagging with joint training methods of a single DNN. To study hierarchical multi-label music tagging, we introduce the multi-label music instrument classification task, which involves instruments organized in a 2-level hierarchy. The contributions of this paper are as follows.
    First, the task provides a more \textbf{realistic scenario}, where we address real music that is typically polyphonic, multi-instrument and diverse in genre. 
    Second, We categorize and explore various \textbf{joint training methods} for DNN under a framework similar to the categorization of conventional hierarchical methods, to facilitate further exploration of the fusion of deep learning and conventional techniques. 
    Finally, for the effective joint training in the multi-label setting, we propose \textbf{ResAtt} and grouped max pooling (\textbf{GMP}) for applying a residual attention layer or max pooling operations to model the connection between fine- and coarse-level tags. 

\section{Related Work}
\label{sec:related_works}
    In the field of audio, existing studies on hierarchical classification have primarily focused on sound event detection tasks. In \cite{cmu2018first_hierarchy}, the connection between coarse- and fine-level tags is formulated as a grouped summation pooling. The formulation requires fine-level predictions that are normalized by a softmax activation, which focuses on single-label tasks. Zharmagambeto \textit{et al.} combines a DNN with a decision tree \cite{amazon2022softtree}, which encourages the DNN to learn a representation that contains hierarchical information. However, a separate classifier is used in the inference phase, so the classifier is not optimized by hierarchical information.

    Nolasco \textit{et al.} studied hierarchical metric learning  of music instruments \cite{rankloss2022} in the Nsynth dataset \cite{nsynth2017} (single note by single instrument). A polyphonic version of the instrument task was addressed in \cite{medlydb2021hierarchy} under the few-shot learning setting, with single-instrument data taken from the stem tracks of MedleyDB dataset \cite{medleydb2014}. After filtering out tags that are not sufficiently fine, \textit{e.g.}, ``Drum", hierarchies are induced  on the basis of instrument categorization used in the music world \cite{hornbostel-sachs}. This idea inspired us to introduce our own tone-base hierarchy (shown in Fig.~\ref{fig:OpenMIC_label}) into the dataset that is used in our experiments.

    Hierarchical multi-label music tagging is discussed in \cite{krause2022sing} in terms of singing activity detection. Loss items are introduced to build the connection between coarse and fine predictions made by a single DNN to improve the detection performance. A framework to describe various DNN-based hierarchical methods is also introduced in \cite{krause2022sing}, however, many of the studies have been devoted to separate multi-model approaches, rather than training these models jointly.
%
\begin{figure*}[htb]
    \begin{minipage}[t]{0.19\linewidth}
        \centering
        \centerline{\includegraphics[width=3.0cm]{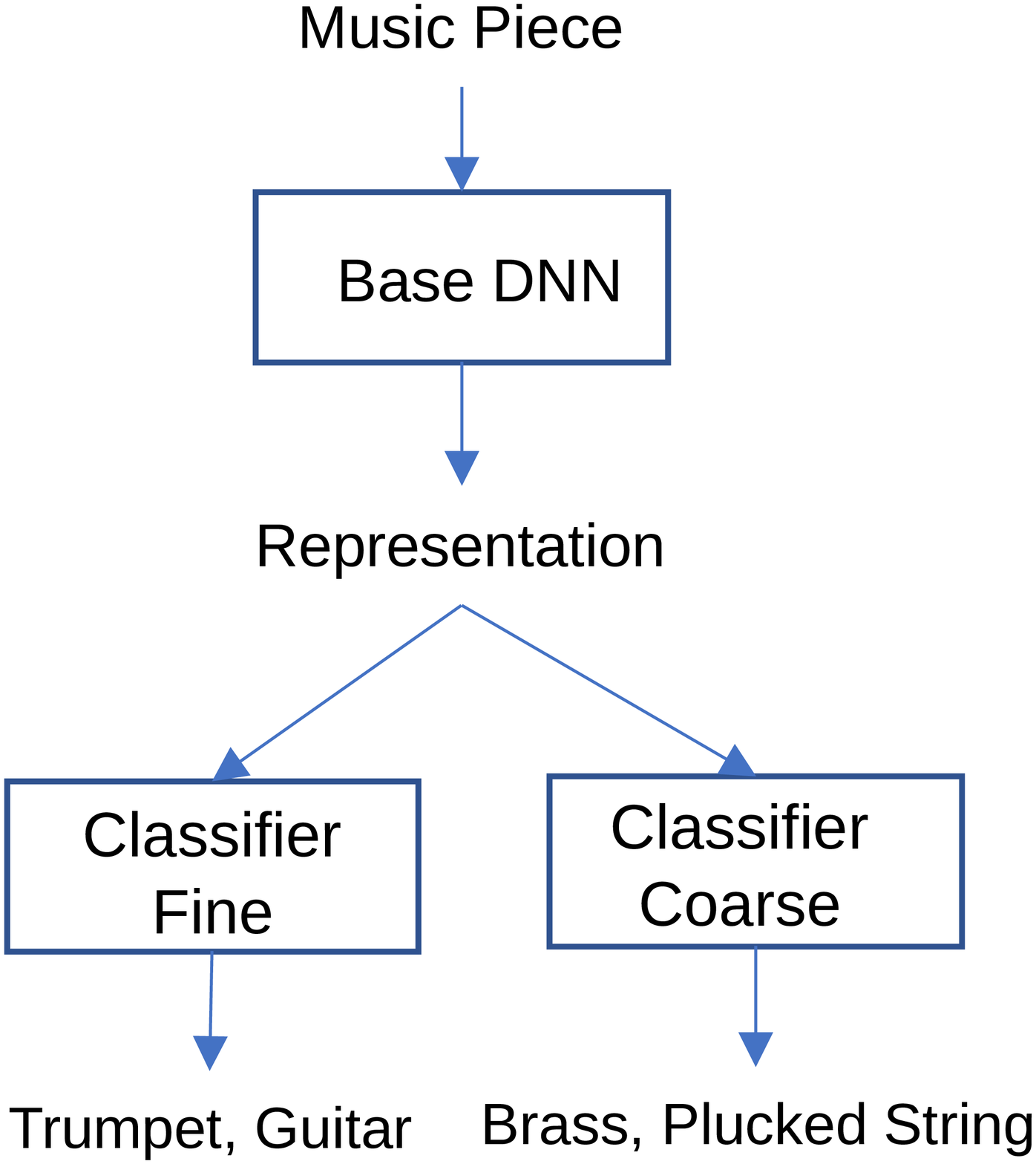}}
        \centerline{(a) Level-wise approach}\medskip
    \end{minipage}
    \begin{minipage}[t]{0.20\linewidth}
        \centering
        \centerline{\includegraphics[width=3.6cm]{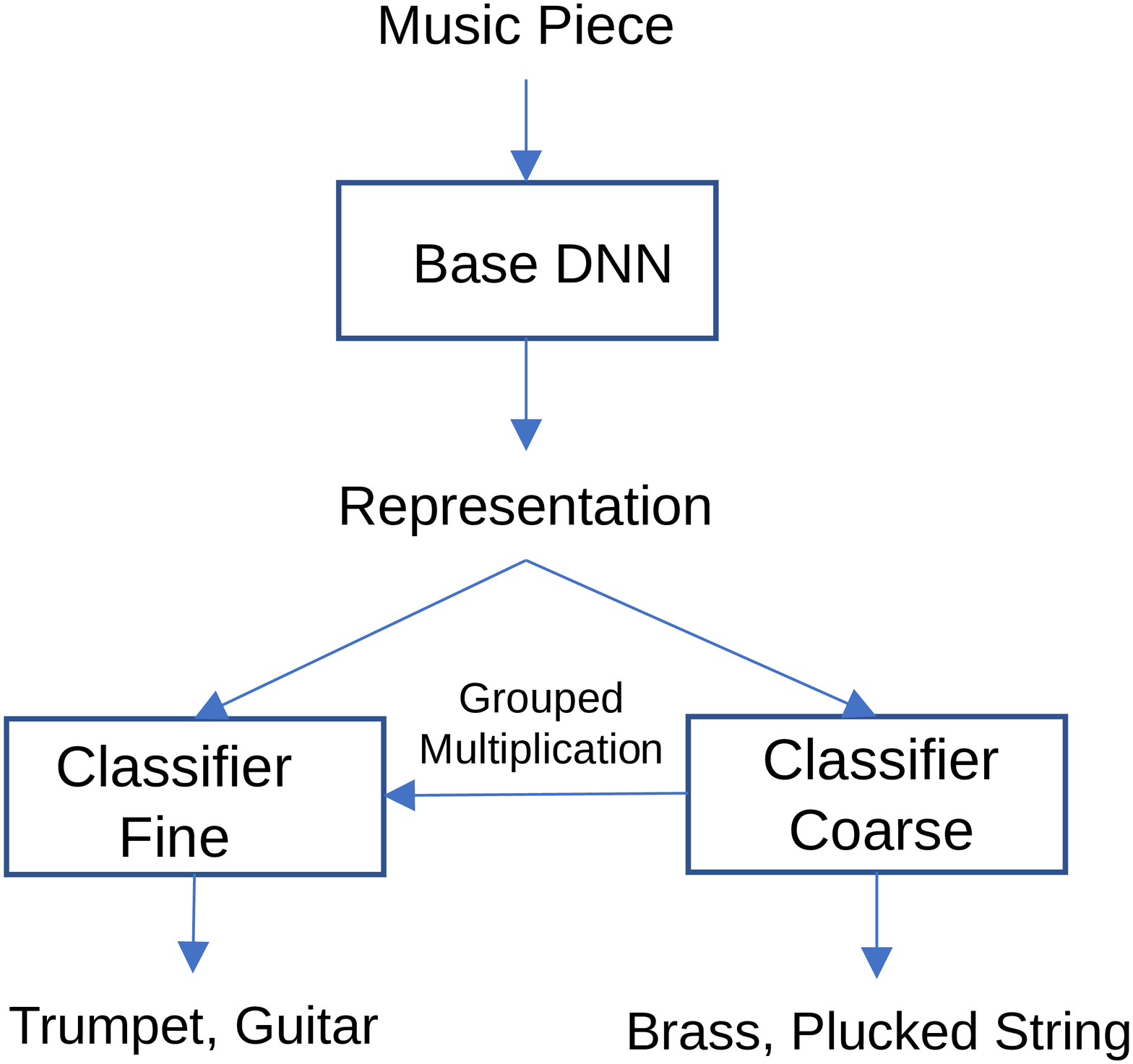}}
        \centerline{(b) Top-down approach}\medskip
    \end{minipage}
    \begin{minipage}[t]{0.24\linewidth}
        \centering
        \centerline{\includegraphics[width=2.52cm]{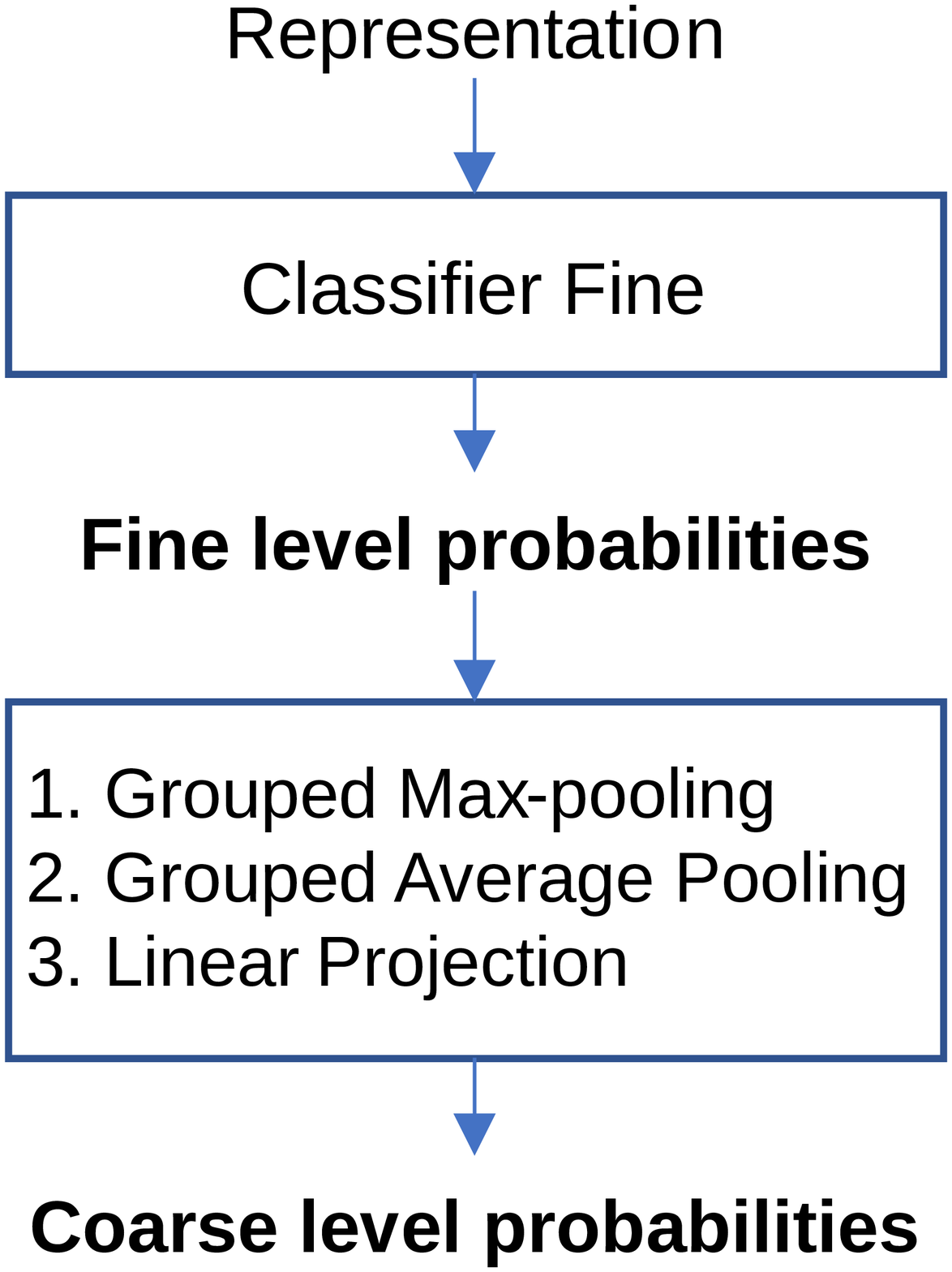}}
        \centerline{(c) Bottom-up approach}\medskip
    \end{minipage}
    \begin{minipage}[t]{0.34\linewidth}
        \centering
        \centerline{\includegraphics[width=5.2cm]{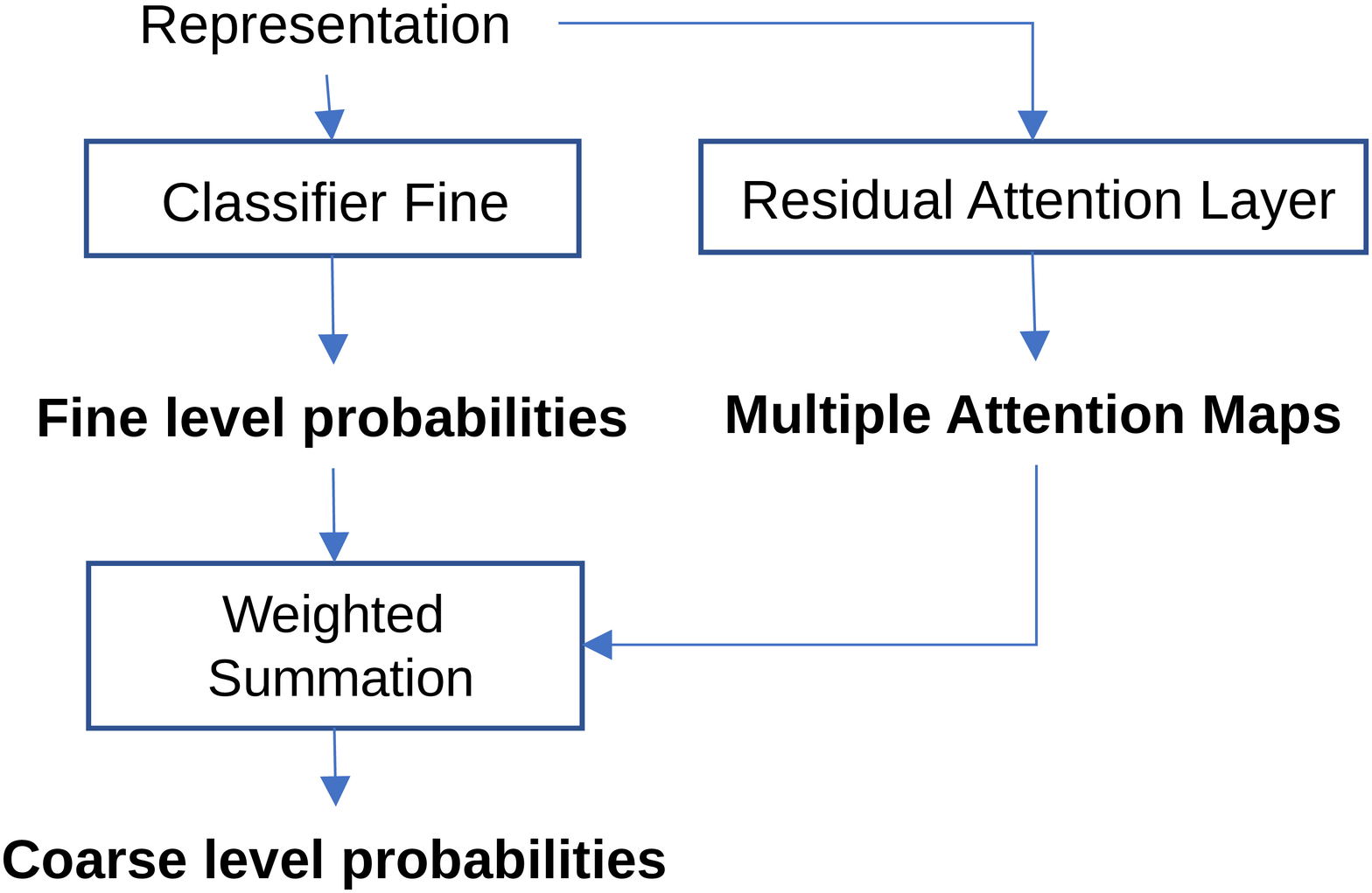}}
        \centerline{(d) ResAtt: Proposed bottom-up approach}\medskip
    \end{minipage}
    \vspace{-3.0mm}
    \caption{Joint training approaches for DNN}
    \label{fig:topology}
    \vspace{-4mm}
\end{figure*}

\section{Proposed Methods}
\label{sec:propose}
    Suppose the $N_\mathrm{dim}$-dimension representation of a music piece is extracted by a base DNN as $\mathbf{Z}^{N_\mathrm{dim}}$. The representation is then projected into fine-level probabilities, $\mathbf{P}_\mathrm{fine} \in \mathbb{R}^{N_\mathrm{fine}\times 1}$, by a classifier through sigmoid activations. Binary classification is performed by converting $\mathbf{P}_\mathrm{fine}$ into binaries using tag-wise thresholds. A 2-level (fine/coarse) hierarchy is assumed. 
\vspace{-2.5mm}
\subsection{Hierarchical Approaches for Joint Training}
\vspace{-1mm}
\label{ssec:topology}
Our main focus in this paper is jointly training a DNN for hierarchical multi-label classification. We begin by summarizing various related methods similarly to the categorization of conventional hierarchical techniques described in \cite{krause2022sing,hierarchical_survey2011}, and demonstrate how deep learning is combined with conventional techniques.

    \noindent\textbf{Level-wise approach}. Conventionally, one model is prepared for each level in the hierarchy \cite{hierarchical_survey2011,krause2022sing}. The idea has been adapted to the joint training framework in \cite{krause2022sing}, which could be illustrated as the structure shown in Fig.~\ref{fig:topology} (a). Although this approach helps the base DNN to learn hierarchical information, classifiers remain independent during training. In \cite{krause2022sing}, two loss items are introduced to optimize these classifiers with hierarchical information.
    
    \noindent\textbf{Top-down approach}. In convention, a model is first trained to classify coarse tags, then separate models are prepared at each coarse tag to classify its child (fine-level) tags \cite{hierarchical_survey2011, FMAhierarchy2020,krause2022sing}. We summarize it as ``coarse first, fine last''. In order to adapt this approach to jointly train a DNN, we connect a simplified soft decision tree (SDT) \cite{SDT2021} similar to \cite{amazon2022softtree} after the base DNN. In an SDT, the fine-level probability is the multiplication of the leaf node and its parent node (coarse-level probability) \cite{SDT2021}, which is called the grouped multiplication between a coarse tag and its child fine-level tags in Fig.~\ref{fig:topology} (b). Since a multi-label task is assumed, softmax activations in the SDT are replaced by sigmoid. Unlike \cite{amazon2022softtree}, we use the SDT for inference, as the SDT is jointly optimized during training. However, in the top-down approach, an error made in the coarse level is difficult to correct, which may affect its fine-level performance \cite{krause2022sing}.

    \noindent\textbf{Bottom-up approach}. In the conventional bottom-up approach, only a non-hierarchical (flat) model is trained with fine-level tags. During inference, coarse tags are predicted by aggregating fine-level predictions with pre-defined rules \cite{krause2022sing,hierarchical_survey2011}. We summarize it as ``fine first, coarse last''. The core problem of the conventional bottom-up approach is that, models are not jointly optimized with ``bottom-up rules", resulting in non-optimal performance. Another major drawback is similar to the top-down approach, where fine-level errors propagate to the coarse level \cite{krause2022sing,hierarchical_survey2011}.
    
    A common rule for the bottom-up aggregation is that, when a fine-level tag is assigned to a music sample, the parent coarse-level tag will be assigned to the music as well, \textit{e.g.}, a music sample will be annotated with ``Woodwind'' if it is annotated with ``Flute''. We call this rule as the grouped max-pooling (GMP), because the rule is equal to applying max-pooling to different groups of child fine-level tags. Obviously, the fine- and coarse-level tags in a hierarchical dataset also satisfy the same rule \cite{medlydb2021hierarchy,krause2022sing,fma_dataset}.
    
    We emphasize the concept of the \textbf{bottom-up approach with joint training}, where DNNs are optimized jointly with aggregation rules. Based on the fact that the fine- and coarse-level labels in a dataset satisfy rules similar to max-pooling, we propose to apply GMP \textbf{during training} (Fig.~\ref{fig:topology} (c)) to inform the DNN of the hierarchical structure in the dataset and improve tagging performance.
\vspace{-5.5mm}
\subsection{ResAtt: Attention-based Bottom-up Method}
\vspace{-0.5mm}
\label{ssec:proposed}
    We also propose \textit{\textbf{ResAtt}}, which models the connection between fine- and coarse-level tags by the attention mechanism. ResAtt can use the attention map with elements valued between 0 - 100\% to tell the system which fine-level tags are and are not important for a specific coarse-level tag. This idea is inspired by the insight that max-pooling is equal to an operation which produces a 0/1 binary attention map to select out the most important element from an input vector. Hence, both proposed methods can be understood as attention-based approaches.
    
    As shown in Fig.~\ref{fig:topology} (d), a residual attention layer is used to generate the attention map for each coarse-level tag, denoted as $\mathbf{W}\in \mathbb{R}^{N_\mathrm{fine}\times N_\mathrm{coarse}}$; the attention map is applied to fine-level predictions, to finally predict coarse-level tags through the following formula:
    \begin{equation}
        \label{eq:resatt}
        \mathbf{P}_\mathrm{coarse} = \mathbf{W}^T \cdot \mathbf{P}_\mathrm{fine} \in \mathbb{R}^{N_\mathrm{coarse}\times 1},
    \end{equation}
    where $T$ is the transpose of matrix. A softmax activation is applied to the $N_\mathrm{fine}$ dimension of attention map $\mathbf{W}$ to ensure that the resulting coarse-level probabilities remain meaningful (below 100\%).

    In ResAtt, to obtain the prediction of coarse-level tags, high quality results from the fine-level classifier are required first. This helps to optimize the classifiers and base DNN jointly during training. Unlike the top-down or the joint training method with GMP, which is informed of the hierarchical structure of the dataset via the tree structure or the max-pooling rule,  ResAtt has no access to such prior knowledge, but explores proper aggregation rules in a data-driven manner. In Sec.~\ref{sec:result}, we will discuss how this feature helps to prevent fine-level errors from propagating to the coarse level.

    Following \cite{cmu2018first_hierarchy, medlydb2021hierarchy}, the overall BCE loss is formulated as the weighted summation of level-wise losses, \textit{i.e.},
    \begin{equation}
    \label{eq:loss_function}
        L_\mathrm{BCE} = \lambda L_\mathrm{BCE}^\mathrm{fine} + (1 - \lambda) L_\mathrm{BCE}^\mathrm{coarse},
    \end{equation}
    where $\lambda$ is the weight for fine level tags.
\section{Evaluation}
\label{sec:eval}
\subsection{Tone-base Hierarchical Dataset}
\label{ssec:MICdataset}
    OpenMIC \cite{humphrey2018openmic} is a music instrument classification dataset that offers a task close to real applications; 10-second music clips of various genres are taken from the FMA dataset \cite{fma_dataset} and annotated concerning 20 instruments in a multi-label manner. While official annotations only include the fine level, we introduce a 2-level instrument hierarchy (Fig.~\ref{fig:OpenMIC_label}) based on the tonal properties of instruments. This pre-processing is similar to the methodology described in \cite{medlydb2021hierarchy}. However, our hierarchy is different from \cite{medlydb2021hierarchy} in that, we include annotations that are considered not ``fine'' enough in \cite{medlydb2021hierarchy} into our hierarchy, such as ``Mallet Percussion" or ``Drums", so we can use all music clips provided by OpenMIC. 

    Since there is no official validation set for hyperparameter tuning, 15\% of the training set data is taken out for validation, following the practice in \cite{gururani2019attentionmic}.  Stratified sampling by scikit-multilearn library \cite{skmultilearn} is used. 
    
    Many instruments remain not annotated in the OpenMIC dataset, so the dataset is released with a masking file, telling users what specific instruments are examined in a track.  We follow the common practice of using this masking file to calculate the loss during training and the metrics during evaluation \cite{gururani2019attentionmic,koutini2022patchout}. The pseudocode for loss calculation can be written as $\mathrm{Loss} = L_\mathrm{BCE}(\mathrm{predictions[mask]}, \mathrm{label[mask]})$.
%
\begin{table*}[htb]
\caption{Results at the tone-base hierarchy (\%)}
\label{tab:result_tone}
\centering
\resizebox{\linewidth}{!}{
\begin{tabular}{clcccccc}
\toprule
    &&\multicolumn{3}{c}{Fine Level}&\multicolumn{3}{c}{Coarse Level}\\
\midrule
{Comments}&{Method}&{ROC-AUC}&{PR-AUC}&{F1 Score}&{ROC-AUC}&{PR-AUC}&{F1 Score}\\
\midrule
{}&{PaSST-S \cite{koutini2022patchout}}&{-}&{84.3}&{-}&{n/a}&{n/a}&{n/a}\\
\hline
{baseline}&{Flat fine-level classification \& inference phase bottom-up}&{91.7}&{85.6}&{83.3}&{92.4}&{89.6}&{84.4}\\
\hline
{Fig.~\ref{fig:topology} (a)}&{Level-wise approach with loss items in \cite{krause2022sing} (our tuning)}&{91.9}&{85.8}&{\textbf{83.7}}&{93.0}&{90.4}&{85.2}\\
\midrule
{Fig.~\ref{fig:topology} (b)}&{Top-down approach}&{91.9}&{85.4}&{83.6}&{93.0}&{90.4}&{85.3}\\
\midrule
&{Bottom-up approach with joint training}\\
{Fig.~\ref{fig:topology} (c)}&{\ \ \ \ 1. Grouped average pooling (GAP)}&{91.7}&{85.6}&{83.2}&{91.7}&{87.6}&{84.6}\\
{Fig.~\ref{fig:topology} (c)}&{\ \ \ \ 2. Linear projection (LP)}&{91.8}&{85.8}&{83.6}&{75.3}&{70.9}&{67.6}\\
{Fig.~\ref{fig:topology} (c)}&{\ \ \ \ 3. Grouped max-pooling (GMP) (proposed)}&{{91.9}}&{{85.9}}&{83.6}&{\textbf{93.1}}&{\textbf{90.7}}&{85.4}\\
{Fig.~\ref{fig:topology} (d)}&{\ \ \ \ 4. ResAtt (proposed)}&{{\textbf{92.0}}}&{\textbf{86.0}}&{\textbf{83.7}}&{\textbf{93.1}}&{\textbf{90.7}}&{\textbf{85.5}}\\
\bottomrule
\end{tabular}}
\vspace{-6mm}
\end{table*}
\subsection{Experiments}
\label{ssec:exp}
    We use the CNN14 architecture pretrained on AudioSet \cite{kong2020panns} as the base DNN, where $N_\mathrm{dim}=2048$, and the fine level classifier is a single linear layer. Raw music data are converted to mono-channel at 16kHz sampling rate, which are further transformed into 64-bin mel-spectrograms (frequency range: [50Hz, 8kHz]), via short-time Fourier transform with 32-ms Hann window and 10-ms hop size. The input audio length is the same as the clip length in OpenMIC.

    We evaluate various approaches to compare them with our proposed methods, of which the model size has been kept in almost the same level. We train the following methods with eight different random seeds and compute their performance metrics. 
    \noindent\textbf{Flat baselines}. We train a CNN14 model as the flat baseline for the fine level. Coarse-level predictions of the model is produced by applying GMP at the inference phase (conventional bottom-up approach used in \cite{amazon2022softtree,krause2022sing}). Fine-level results reported in \cite{koutini2022patchout} are compared with our baseline. 
    \noindent\textbf{Level-wise approach}. We evaluate the DNN shown in Fig.~\ref{fig:topology} (a) with two parallel linear layers. We introduce loss items proposed in \cite{krause2022sing} into our architecture and train the DNN with the BCE loss in Eq.~\ref{eq:loss_function}. 
    \noindent\textbf{Top-down approach}. We evaluate the DNN with an SDT classifier shown in Fig.~\ref{fig:topology} (b), where each level in the tree is a single linear layer. 
    \noindent\textbf{Bottom-up approach with joint training}. We evaluate ResAtt in Fig.~\ref{fig:topology} (d), whose attention layer is implemented by reshaping the output of a linear layer, which is as light weighted as the level-wise or the top-down approach. We evaluate the proposed joint training method with GMP as well. To compare with ResAtt, we replace the attention mechanism with a linear projection (LP) layer. Similarly, we replace max-pooling with average pooling to form the grouped average pooling (GAP) method to compare with the joint training method with GMP. We jointly train GAP and LP as in Fig.~\ref{fig:topology} (c).

    During the training, the batch size is 16. The maximum learning rate is 1e-4. The first 5 epochs are trained with linearly increasing learning rates (linear warm-up) to make the training more stable. The Adam optimizer \cite{adam_optimizer} with weight decay of 1e-4 is used. We utilize SpecAugment \cite{specaugment2019} for data augmentation. Grid searches are carried out for the weight $\lambda$ in the loss function (Eq.~\ref{eq:loss_function}), where $\lambda \in [0.70,0.75,0.80,0.85,0.90]$. The setting with the lowest validation loss for fine-level tags is used for evaluation.
\subsection{Metrics}
\label{ssec:metric}
    For objective evaluation, we use the macro average of ROC-AUC, PR-AUC (also known as mean average precision), and F1 score. All metrics are calculated using the scikit-learn library \cite{scikit-learn}. F1 scores are calculated on the basis of 0/1 binary predictions, which are produced by thresholds optimized on the validation set.
\section{Results}
\label{sec:result}
The average of evaluation results across various random seeds are presented in Tab.~\ref{tab:result_tone}, where digits denote various design choices. 

Our model for fine-level classification outperformed PaSST-S in \cite{koutini2022patchout}. This implies that our choice of base DNN as well as our training settings are feasible. 

The level-wise approach combined with loss items in \cite{krause2022sing} improved coarse-level performance by a large margin, and improved the fine-level performance especially for the F1 Score. However, it is difficult to interpret the decision procedures within this model. Meanwhile, the top-down approach is interpretable because of its tree structure in the classifier \cite{SDT2021}. In the coarse level, the top-down approach has a performance similar to the level-wise approach, but in the fine level, the approach resulted in deteriorated PR-AUC value. As described in Sec.~\ref{ssec:topology}, its fine-level performance may have been limited by its coarse-level performance.

The only difference between the proposed joint training method with GMP and our baseline is that, the proposed method optimizes the DNN with the GMP rule by joint training. The joint training method with GMP is interpretable through the GMP rule, \textit{e.g.}, ``Woodwind'' is predicted because at least one of ``Flute'', ``Clarinet'' or ``Saxophone'' is predicted with high probability. As mentioned in Sec.~\ref{ssec:topology}, fine- and coarse-level  annotations in the tone-base hierarchical dataset satisfy rules similar to GMP, which explains the large performance improvements brought by GMP joint training. Replacing the max-pooling with average pooling operations results in the GAP method. Since in the dataset a coarse-level label is not the average of its child fine-level labels, jointly training the DNN with GAP resulted in lower performance.

\begin{figure}[tb]
    \centering
    \includegraphics[width= \linewidth]{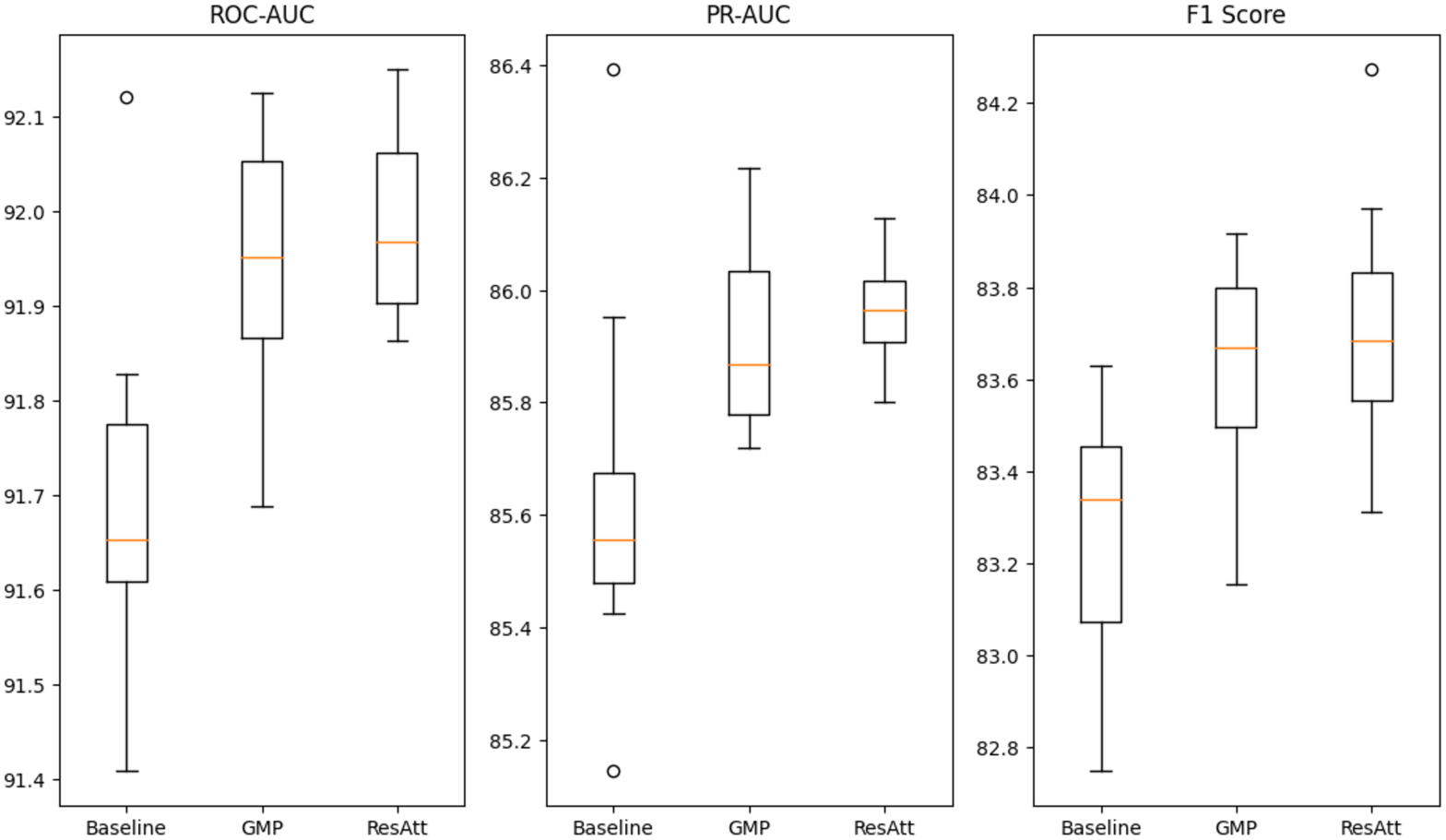} 
    \vspace{-7mm}
    \caption{The boxplot of fine-level metrics for the baseline and proposed models. Orange lines denote the medians}
    \label{fig:boxplot}
    \vspace{-5.5mm}
\end{figure}
\begin{figure}[tb]
    \includegraphics[width= \linewidth]{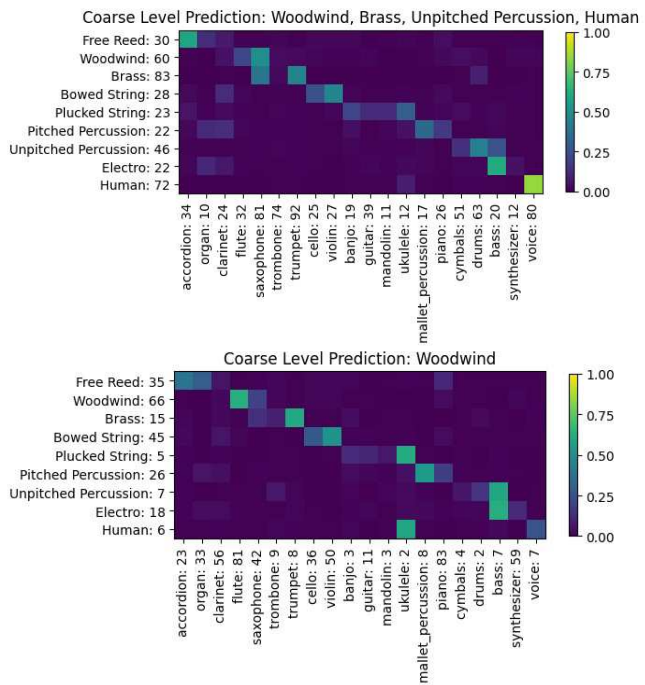} %
    \vspace{-8.25mm}
    \caption{Attention maps. Coarse tag predictions are listed at the top; tag-wise probabilities (\%) are listed with tag names.}
    \label{fig:attention_map}
    \vspace{-5mm}
\end{figure}

Contrary to the joint training method with GMP, ResAtt is unaware of the hierarchical structure in the dataset during training, but attempts to extract the aggregation rule from labels alone. ResAtt achieves scores that are equal to or slightly higher than the joint training method with GMP. The LP method utilizes a learnable projection layer, instead of the attention layer, but failed to predict coarse-level tags. This is because that a linear layer cannot flexibly adjust its projection rule for various input, which reveals how important it is to use attention mechanism in the bottom-up approach with joint training.

Since the improvements from the baseline in fine-level metrics are small in terms of absolute value, we compare the proposed methods with the baseline in box plots (Fig. \ref{fig:boxplot}). Higher medians and boxes demonstrate their advantages over the baseline. Although ResAtt has slightly higher medians, the difference between the joint training method with GMP and ResAtt is not significant.

Attention map samples with tag-wise probabilities made by ResAtt are shown in Fig.~\ref{fig:attention_map}. All maps are similar to the GMP operation, but vary depending on input music, which implies that ResAtt has extracted the aggregation rule without prior knowledge. On the lower side, although the ``piano'' in the fine level is wrongly predicted as 83\%, ResAtt did not pass this error to the coarse level by giving ``piano'' less attention. This shows that ResAtt can learn flexible aggregation that is difficult to describe with fixed rules, and in some cases, it can even prevent the error from propagating to the coarse level. Such behaviors are impossible for the joint training method with GMP, as the method accumulates errors made in the fine level \cite{krause2022sing,hierarchical_survey2011}, which is a possible reason for the slightly higher F1 Score of ResAtt in the coarse level. The decision procedure within ResAtt is interpretable on the basis of attention map visualization.
\section{ Conclusion and Future Work}
\vspace{-1.5mm}
\label{sec:conclusion}
We investigated hierarchical multi-label music instrument classification as a case study of hierarchical music tagging. We extended hierarchical instrument classification to the multi-label setting and realistic music data, with an induced tone-base hierarchy. Various hierarchical methods that jointly train a DNN are summarized in the context of the fusion of deep learning and conventional techniques. For the effective joint training in the multi-label setting, we propose two methods to model the connection between fine- and coarse-level tags, where one uses the attention mechanism obtained in a data-driven manner, the other uses rule-based grouped max-pooling, which is explained as a binary attention. Evaluation results indicate that proposed methods, especially the ResAtt, are promising as the bottom-up methods with joint training. In addition to its performance, ResAtt can learn flexible aggregation that is difficult to describe with fixed rules and even prevent errors from propagating. By visualizing attention maps, the interpretability of ResAtt can be enhanced. Future work involves extending current methods to other tasks in music tagging.

\vfill\pagebreak

\bibliographystyle{IEEEbib}
\bibliography{strings}
    
\end{document}